\documentstyle[12pt,a41,axodraw]{article}

\begin{document}

\thispagestyle{empty}

\begin{flushleft}
DESY  98-176 \hfill
{\tt hep-ph/9811351}\\
November 1998
\end{flushleft}

\setcounter{page}{0}

\mbox{}
\vspace*{\fill}
\begin{center}
{\LARGE\bf Heavy flavour contributions to the}

\vspace{2mm}
{\LARGE\bf deep inelastic scattering sum rules}

\vspace*{20mm}
\large
{J. Bl\"umlein and W.L. van Neerven 
\footnote{On leave of absence from
Instituut-Lorentz, University of Leiden,P.O. Box 9506, 2300 RA Leiden,\\
The Netherlands
.}}
\\

\vspace{2em}

\normalsize
{\it DESY-Zeuthen, Platanenallee 6, D-15738 Zeuthen, Germany}%

\vspace*{\fill}
\end{center}
\begin{abstract}
\noindent
We have calculated the first and second order corrections to several 
deep inelastic sum rules which are due to heavy flavour contributions.
A comparison is made with the existing perturbation series which has
been computed up to third order for massless quarks only. In general it 
turns out that the effects of heavy quarks are very small except when
$Q \sim m$ or $Q \gg m$. Here $Q$ and $m$ denote the virtual mass of the
vector boson and the mass of the heavy quark, respectively. For $Q \gg m$
the radiative corrections reveal large logarithms of the type 
$\ln Q^2/m^2$ which have to be absorbed in the running coupling constant 
so that the number of light flavours $n_f$ is enhanced by one unit. 
However this has to happen at much larger values of Q i.e. $Q \sim 6.5~ m$ 
than one usually assumes for the flavour thresholds which appear in the 
running coupling constant. An alternative description for the heavy
flavour dependence of the running coupling constant in the context
of the MOM-scheme is discussed.
\end{abstract}
\vspace*{\fill}

\newpage\noindent
The study of QCD sum rules, as represented by the first moments of the
deep inelastic structure functions, has lead to a deeper insight of
the behaviour of the perturbation series. This became possible after
new techniques were invented to evaluate the Feynman integrals up to
four-loop order. Examples of these techniques are infrared rearrangement
\cite{ckt}, 
integration by parts \cite{chtk}, and the $R^*$-operation \cite{chsm}.
Also important was the appearance of new algebraic manipulation programs 
like {\tt FORM} \cite{form} which enables us to evaluate the complicated
traces of the huge amount of Feynman graphs characteristic of higher
order loop calculations. At this moment the sum rules computed up to 
third order in $\alpha_s$ are represented by the first Bj{\o}rken 
(polarized)
sum rule \cite{bjork1}, the second Bj{\o}rken (unpolarized) sum rule
\cite{bjork2} and the Gross-Llewellyn Smith sum rule \cite{grls}. The
perturbation series for these sum rules show a similar behaviour as is
observed for other quantities which are calculated up to third order
like e.g. the $Z$-boson and $\tau$-lepton decay widths
(for a review of the literature see \cite{kast}). Quantities
computed up to a very high order in perturbation theory provide us
with a very good tool to understand methods used in improved
perturbation theory. Examples are the principle of minimal sensitivity
(PMS \cite{stev}) end the effective charge approach (ECH \cite{grun}). 
These methods were applied \cite{kast} to the above sum rules to obtain
an estimate of the unknown order $\alpha_s^4$ contribution. Another way
to get the latter term is to use Pade-approximants as carried out
in \cite{sael} (for an estimate using renormalons see also \cite{elga}).
One of the remarkable results of these methods is that all estimates 
agree
very well with each other. Apart from the theoretical interest there is
also a practical one. Quanities which can be calculated up to a very high
order in perturbation theory provide us with an excellent tool to measure
the running coupling constant $\alpha_s$. Notice that in many cases the 
perturbation series is only known up to next-to-leading order (NLO) which
means that, apart from some resummation of dominant terms, we have no
control on the higher order corrections. An example of the determination 
of $\alpha_s$ is given in \cite{elka} where it is extracted via the 
polarized Bj{\o}rken sum rule from the data obtained for the longitudinal
structure function $g_1(x,Q^2)$. 

The order $\alpha_s^3$ corrections to the sum rules mentioned above
have been carried out in \cite{latk} (the unpolarized Bj{\o}rken sum 
rule) and
\cite{lave} (the polarized Bj{\o}rken sum rule and the Gross-Llewellyn 
Smith 
sum rule). In these calculations only massless quarks were considered
but mass effects coming from the contribution of heavy quarks were 
omitted.
The latter are important because apart from additional corrections the
mass effects indicate when a heavy quark has to be treated as a massless
or as a massive quark. This also indicates which number
of light flavours $n_f$ has to be chosen in the perturbation series in
particular for the running coupling constant at a given value of $Q^2$. 
Here $Q$ denotes the virtual mass of the intermediate vector boson in 
deep inelastic lepton hadron scattering.
Before presenting the heavy flavour contributions we first give the
definitions of the three aforementioned sum rules and the corresponding
perturbation series corrected up to third order in $\alpha_s$.
The polarized \cite{bjork1} and unpolarized Bj{\o}rken \cite{bjork2} sum 
rules
are defined by
\begin{eqnarray}
\label{eq1}
\Delta g_1(Q^2)\equiv \int_0^1\, dx\, \left [g_1^{ep}(x,Q^2) - 
g_1^{en}(x,Q^2) 
\right ] = \frac{1}{6} \left| \frac{G_A}{G_V} \right| A^{g_1}(Q^2) \,,
\end{eqnarray}
and
\begin{eqnarray}
\label{eq2}
 \Delta F_1(Q^2)& \equiv& \int_0^1\, dx\, \left [F_1^{\bar \nu p}(x,Q^2) 
- F_1^{\nu p}(x,Q^2) \right ] = K(n_f) \, A^{F_1}(n_f,Q^2)\,
\nonumber\\[2ex]
 K(3) &=& 1 + \sin^2 \theta_c \quad ( SU_F(3)) 
\qquad K(4) = 1 \quad (SU_F(4)) \,,
\end{eqnarray}
respectively, whereas the Gross-Llewellyn Smith sum rule \cite{grls} is
given by
\begin{eqnarray}
\label{eq3}
\Delta F_3(Q^2)&\equiv& \int_0^1\, dx\, \left [F_3^{\bar \nu p}(x,Q^2) 
+ F_3^{\nu p}(x,Q^2) \right ] = K(n_f) \, A^{F_3}(n_f,Q^2)\,
\nonumber\\[2ex]
 K(3) &=& 6 - 2 \sin^2 \theta_c \quad ( SU_F(3)) \qquad K(4) = 6 \quad 
(SU_F(4)) \,.
\end{eqnarray}
Here $\theta_c$ denotes the Cabibbo angle and for the constant $K(n_f)$ 
we have 
quoted the values given by the flavour group $SU_F(n_f)$ for $n_f=3,4$, 
where
$n_f$ represents the number of light flavours. 
\begin{table}
\begin{center}
\begin{tabular}{|c|c|c|c|}\hline
\multicolumn{4}{|c|}{$A(Q^2)$} \\ \hline \hline
& $A^{ g_1}(Q^2)$& $A^{ F_1}(Q^2)$& $A^{ F_3}(Q^2)$\\ \hline
\hline
           &                &                    &             \\
$a_1$ &  -1        &  -$\frac{2}{3}$     &  -1           \\[1mm]
$a_2$ &  -$\frac{55}{12}$  & -$\frac{23}{6}$ & -$\frac{55}{12}$  \\[2mm]
$b_2$ &  $\frac{1}{3}$     &  $\frac{8}{27}$     &  $\frac{1}{3}$ \\[2mm] 
$a_3$ & -$\frac{13841}{216}$-$\frac{44}{9}\zeta(3)$ + $\frac{55}{2}\zeta(5)$ &
 -$\frac{4075}{108}$+$\frac{622}{27}\zeta(3)$ - $\frac{680}{27}\zeta(5)$ & 
 -$\frac{13841}{216}$-$\frac{44}{9}\zeta(3)$ + $\frac{55}{2}\zeta(5)$\\[2mm]
$b_3$ & $\frac{10339}{1296}$+$\frac{61}{54}\zeta(3)$ - $\frac{5}{3}\zeta(5)$ & 
  $\frac{3565}{648}$-$\frac{59}{27}\zeta(3)$ + $\frac{10}{3}\zeta(5)$     &
  $\frac{10009}{1296}$+$\frac{91}{54}\zeta(3)$ - $\frac{5}{3}\zeta(5)$\\[2mm]
$c_3$ & -$\frac{115}{648}$  & -$\frac{155}{972}$ & -$\frac{115}{648}$ \\[2mm]
$a_4^{\rm PMS}(3)$ &  - 130  & -133 & -130 \\[2mm] \hline
\end{tabular}
\end{center}
\caption{\sf
The coefficients in the ${\overline {MS}}$-scheme of the perturbation series 
(\ref{eq4}) corresponding to the three sum rules in Eqs. 
(\ref{eq1})-(\ref{eq3}).}
\label{tab1}
\end{table}
The perturbation series of the above sum rules in the case of  massless 
quarks 
can be written up to third order in $\alpha_s$ as
\begin{eqnarray}
\label{eq4}
A^{\rm r}(n_f,Q^2)&=&  1 + \frac{\alpha_s(n_f,\mu^2)}{\pi} a_1 +
\left (\frac{\alpha_s(n_f,\mu^2)}{\pi} \right )^2 \left [
-a_1 \beta_0(n_f) \ln\left (\frac{Q^2}{\mu^2} \right ) + a_2 + b_2n_f
\right ] 
\nonumber\\[2ex]
&&  + \left (\frac{\alpha_s(n_f,\mu^2)}{\pi} \right )^3 \left [
a_1 \beta_0^2(n_f) \ln^2 \left (\frac{Q^2}{\mu^2} \right )
- \Big \{ a_1 \beta_1(n_f) + 2 \beta_0(n_f) \Big ( a_2 + b_2n_f \Big ) 
\Big \} \right.
\nonumber\\[2ex]
&& \left. \times \ln\left (\frac{Q^2}{\mu^2} \right )  
 +  a_3 + b_3n_f + c_3n_f^2 \right ] \quad \mbox{with} \quad 
r=g_1,F_1,F_3 \,,
\end{eqnarray}
where $\beta_0$ and $\beta_1$ stand for the first and second order
contributions to the $\beta$-function which are given by
\begin{eqnarray}
\label{eq5}
\beta_0(n_f)=\frac{11}{4}-\frac{1}{6}n_f \qquad 
\beta_1(n_f)=\frac{51}{8}-
\frac{19}{24}n_f~.
\end{eqnarray}
The other coefficients $a_i, b_i, c_i$, which are computed in the
${\overline {\rm MS}}$-scheme in \cite{latk} and \cite{lave}, are given 
in 
table \ref{tab1}.
As has been mentioned above, the order $\alpha_s^4$ contribution to Eq.
(\ref{eq4}) is not known. However, there exist  some estimates. Here we
will adopt the results obtained from PMS given in \cite{kast}. They will
be denoted by
\begin{eqnarray}
\label{eq6}
\delta A^{\rm r, PMS}(Q^2) = \left (\frac{\alpha_s(n_f,\mu^2)}{\pi} 
\right )^4 
\left [ a_4^{\rm PMS} (n_f) \right ]~,
\end{eqnarray}
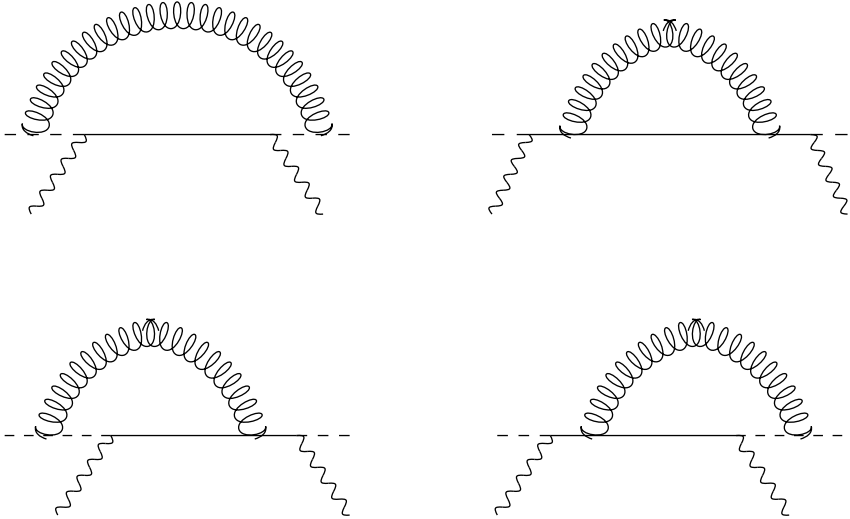
\begin{figure}
\begin{center}
  \begin{picture}(150,100)(0,0)
  \DashLine(10,30)(40,30){4}
  \DashLine(110,30)(140,30){4}
  \Line(40,30)(110,30)
  \Photon(20,0)(40,30){2}{5}
  \Photon(130,0)(110,30){2}{5}
  \GlueArc(75,20)(55,170,10){5}{30}
\end{picture}
\hspace*{1cm}
  \begin{picture}(150,100)(0,0)
  \DashLine(8,30)(22,30){4}
  \DashLine(128,30)(142,30){4}
  \Line(22,30)(128,30)
  \Photon(8,0)(22,30){2}{5}
  \Photon(142,0)(128,30){2}{5}
  \GlueArc(87,20)(50,170,101){5}{11}
  \GlueArc(63,20)(50,79,10){5}{11}
\end{picture}
\end{center}
\begin{center}
  \begin{picture}(150,100)(0,0)
  \DashLine(10,30)(50,30){4}
  \DashLine(120,30)(140,30){4}
  \Line(50,30)(120,30)
  \Photon(30,0)(50,30){2}{5}
  \Photon(140,0)(120,30){2}{5}
  \GlueArc(75,20)(50,170,98){5}{11}
  \GlueArc(55,20)(50,82,10){5}{11}
\end{picture}
\hspace*{1cm}
  \begin{picture}(150,100)(0,0)
  \DashLine(10,30)(30,30){4}
  \DashLine(100,30)(140,30){4}
  \Line(30,30)(100,30)
  \Photon(10,0)(30,30){2}{5}
  \Photon(120,0)(100,30){2}{5}
  \GlueArc(95,20)(50,170,98){5}{11}
  \GlueArc(75,20)(50,82,10){5}{11}
\end{picture}
\caption[]{\sf Forward Compton scattering graphs for heavy flavour
production:
 $W + d \rightarrow g + c$ . The down quark and the charm quark  are
 indicated by a   dashed line and a solid line, respectively.}
\label{fig1}
\end{center}
\end{figure}
\noindent
where the coefficient $a_4^{\rm PMS}(3)$ is given in table \ref{tab1}.
It turns out that the other estimates originating from ECH \cite{kast}
and the Pade-technique \cite{sael} are very close to the PMS value.
Besides the sum rules above we also have the Adler sum rule \cite{adl} 
given by
\begin{eqnarray}
\label{eq7}
\Delta F_2(Q^2)&\equiv&\int_0^1\, \frac{dx}{x} \,\left 
[F_2^{\bar \nu p}(x,Q^2)
- F_2^{\nu p}(x,Q^2) \right ] = K(n_f) \,
\nonumber\\[2ex]
 K(3) &=& 2 + 2 \sin^2 \theta_c \quad (SU_F(3)) \qquad K(4) = 
2\quad (SU_F(4))
\,,
\end{eqnarray}
which holds in all orders of perturbation theory. Furthermore it does 
not receive higher twist contributions or mass corrections. The latter
we have checked in our computations presented below. 
The coefficients in table \ref{tab1} are only determined for massless
quarks (see \cite{latk,lave}). In the subsequent part of the paper we
will show how the perturbation series is modified by including 
mass corrections due to heavy flavour contributions. 

In our calculations
we assume that in addition to the gluon the proton only contains three 
light flavours given by the quarks $u,d,s$,
 including their anti-particles. The
heavy quarks only show up in the final state. Since the sum rules 
presented above only involve non-singlet contributions the
perturbation series for heavy flavour contributions in the case of 
neutral current interactions starts in order $\alpha_s^2$. However, for
the charged current interaction we get already contributions on the Born 
level.  Starting with the latter interaction $\Delta F_1$ in Eq. 
(\ref{eq2}) and $\Delta F_3$ in Eq. (\ref{eq3}) are in lowest order given 
by the flavour excitation process
\begin{eqnarray}
\label{eq8}
d\, (\bar d)\, + \, W \rightarrow c \, (\bar c) \,.
\end{eqnarray}
In the process above we have only considered charm production because
the other heavy quarks are heavily suppressed by the mixing angles 
occurring 
in the Kobayashi-Maskawa matrix. Moreover the relevant values of $Q^2$ 
are so small that they are far below the thresholds of bottom and top 
production.
\begin{figure}
\begin{center}
  \begin{picture}(150,100)(0,0)
  \DashLine(10,30)(140,30){4}
  \Photon(10,0)(30,30){2}{5}
  \Photon(140,0)(120,30){2}{5}
  \GlueArc(75,20)(55,170,120){4}{10}
  \GlueArc(75,20)(55,60,10){4}{10}
  \CArc(75,60)(30,19,161)
  \CArc(75,75)(30,199,341)
\end{picture}
\hspace*{1cm}
  \begin{picture}(150,100)(0,0)
  \DashLine(10,30)(140,30){4}
  \Photon(10,0)(20,30){2}{5}
  \Photon(140,0)(130,30){2}{5}
  \GlueArc(85,20)(55,170,130){4}{8}
  \GlueArc(65,20)(55,50,10){4}{8}
  \CArc(75,55)(28,19,161)
  \CArc(75,70)(28,199,341)
\end{picture}
\end{center}
\begin{center}
  \begin{picture}(150,100)(0,0)
  \DashLine(10,30)(140,30){4}
  \Photon(30,0)(50,30){2}{5}
  \Photon(140,0)(120,30){2}{5}
  \GlueArc(70,20)(50,170,126){4}{8}
  \GlueArc(60,20)(50,54,10){4}{8}
  \CArc(65,55)(28,18,162)
  \CArc(65,70)(28,198,342)
\end{picture}
\hspace*{1cm}
  \begin{picture}(150,100)(0,0)
  \DashLine(10,30)(140,30){4}
  \Photon(10,0)(30,30){2}{5}
  \Photon(120,0)(100,30){2}{5}
  \GlueArc(90,20)(50,170,126){4}{8}
  \GlueArc(80,20)(50,54,10){4}{8}
  \CArc(85,55)(28,18,162)
  \CArc(85,70)(28,198,342)
\end{picture}
\caption[]{\sf
Forward Compton scattering graphs for heavy flavour 
production:
 $V + q \rightarrow q' + H + \bar H$. The light quarks $q$ and the
  heavy quarks $H$ are indicated by   a dashed line and a solid line,
  respectively.}
\label{fig2}
\end{center}
\end{figure}
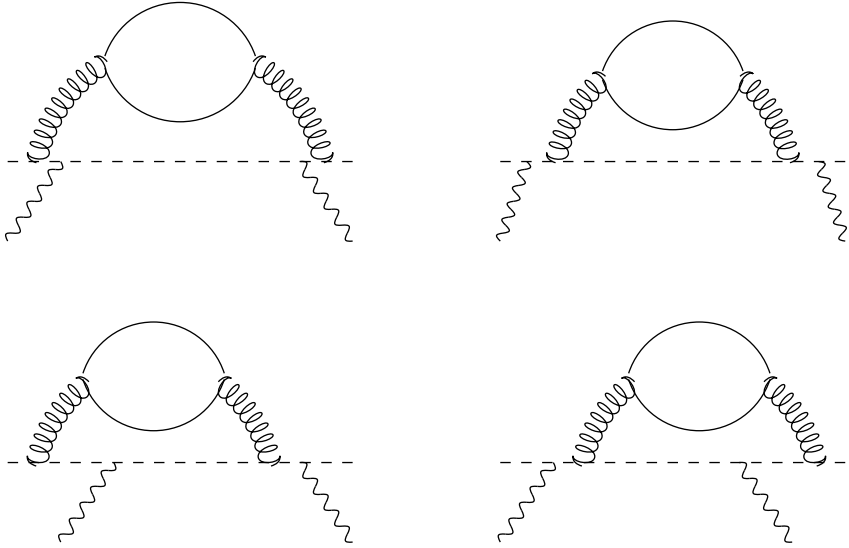
Further notice that the integrals over the strange quark and the
anti-strange quark densities cancel against each other in the computation
of $\Delta F_1$ and $\Delta F_3$ so that the process in Eq. (\ref{eq8}) 
does not contribute to the sum rules  when $d$ is replaced by $s$. In 
NLO one gets contributions from the virtual corrections to reaction 
(\ref{eq6}) and the gluon bremsstrahlungs process (see Fig. \ref{fig1}).
The latter is given by
\begin{eqnarray}
\label{eq9}
d \, (\bar d)\, + \, W \rightarrow c\, (\bar c)\, + \,g~.
\end{eqnarray}
The coefficient functions corresponding to the structure functions $F_i$
($i=1,2,3,L$) have been computed for processes (\ref{eq8}) and 
(\ref{eq9}) in \cite{gott} \footnote{Notice that $A_2$ in table I 
of \cite{gott} should be $K_A$ and not $K_A/2$.} for $m_d=0$ and 
$m_c\not =0$ (see also \cite{gkr}). Notice that due to the non-vanishing 
mass of the charm quark one gets already a contribution to $F_L$ on the 
Born level from reaction (\ref{eq8}). If the quark in the initial state 
becomes massive the above processes also contribute to the neutral 
current interaction where $W$ is repaced by $Z$ or $\gamma$. The
computation of the coefficient functions for the neutral current 
interaction where the masses of the initial and final state quarks are 
equal has been treated in \cite{teve}. Integration of the coefficient 
functions over the scaling variable $x$ provides us with the result for 
the unpolarized Bj{\o}rken sum rule
\begin{eqnarray}
\label{eq10}
A_{\rm H}^{\rm F_1,(1)}(Q^2,m^2)= \left [\frac{1}{1 + \xi}+ 
C_F \frac{\alpha_s(\mu^2)}{4\pi} \left \{
- \frac{1}{1+\xi} - \frac{2}{\xi}  -  \left ( \frac{6}{1 + \xi}
 - \frac{2}{\xi^2}  \right ) \ln \left (1 + \xi \right )
\right \} \right ] \sin^2 \theta_c \,,
\end{eqnarray}
and the Gross-Llewellyn Smith sum rule
\begin{eqnarray}
\label{eq11}
A_{\rm H}^{\rm F_3,(1)}(Q^2,m^2)=\left [- \frac{1}{3(1 + \xi)}+  
C_F \frac{\alpha_s(\mu^2)}{4\pi} \left \{ \frac{1}{1 + \xi}  
+ 2 \frac{\ln(1 + \xi)}{1 + \xi}  \right \} \right ] \sin^2 \theta_c
\qquad  \xi = \frac{Q^2}{m^2} \,,
\end{eqnarray}
respectively, where $C_F$ denotes the colour factor which in QCD reads
$C_F=4/3$. Furthermore, we have also checked that the corrections to the
Adler sum rule in Eq. (\ref{eq7}) are zero as expected.
The above expressions have to be added to the light quark contribution
$A^{\rm r}(3,Q^2)$ in Eq. (\ref{eq4}) to obtain the $O(\alpha_s)$ 
mass corrections to the
sum rules in Eqs. (\ref{eq2}), (\ref{eq3}) with $K=K(4)$. 
In the perturbation series above the following limits are of interest.
When $Q^2$ is much larger than the mass of the charm quark, i.e.
$\xi \rightarrow \infty$, then the corrections in Eqs. (\ref{eq10}) and
(\ref{eq11}) tend to zero. This means that the sum rules are presented
in a four light flavour scheme. However, when the mass of the charm quark
becomes much larger than $Q^2$, i.e. $\xi \rightarrow 0$, then the heavy
quark mass corrections are non-vanishing. After adding them to the
massless quark result $A^{\rm r}(3,Q^2)$ in Eq. (\ref{eq4}) one can 
extract an overall factor $K$ which
turns out to be $K(3)$ which is the value in the three light flavour 
scheme. This is expected because for infinite mass the heavy
flavour disappears from the theory. Unfortunately this does not happen
for the Adler sum rule in Eq. (\ref{eq7}) because it is insensitive
to the mass of the heavy flavours. The next process which shows up
in neutral current as well as in charged current interactions is given
by gluon splitting into a heavy quark anti-quark pair (see Fig.
\ref{fig2})
\begin{eqnarray}
\label{eq12}
q + V \rightarrow q' + Q + \bar Q \quad \mbox{with} \quad V
=\gamma,Z,W \,.
\end{eqnarray}
The coefficient function for $g_1$, which is the same as for $F_3$, has
been calculated in \cite{bmsn}. Notice that the heavy quark loop 
contribution
in  Fig.~\ref{fig2} to the gluon self-energy $\Pi(p^2,m^2)$, where $p$
denotes
the gluon momentum, has been renormalized in such a way that 
$\Pi(0,m^2)=0$.
This implies that heavy quarks are decoupled from the running coupling 
constant.
The result for the polarized Bj{\o}rken sum rule and the Gross-Llewellyn 
Smith 
sum rule becomes equal to
\begin{eqnarray}
\label{eq13}
&& A_{\rm H}^{g_1,(2)}(Q^2,m^2)=  A_{\rm H}^{F_3,(2)}(Q^2,m^2) =
\left (\frac{\alpha_s(n_f,\mu^2)}{4\pi} \right )^2 C_F T_f
\left [ \left ( \frac{1}{105} \xi^2 + \frac{16}{45} \xi \right ) 
\ln \xi \right.
\nonumber\\[2ex]
&& \left. +\frac{1}{\lambda^4}\left ( \frac{2}{105} \xi + 
\frac{2783}{315} 
+ \frac{6740}{63} \frac{1}{\xi} + \frac{137552}{315} \frac{1}{\xi^2} 
+ \frac{62528}{105} \frac{1}{\xi^3} \right )-\frac{1}{\lambda^5} 
\left ( 
\frac{1}{105} \xi^2 + \frac{142}{315} \xi + \frac{494}{63} \right. 
\right.
\nonumber\\[2ex]
&&  \left. \left. + \frac{1516}{21} \frac{1}{\xi} + \frac{23024}{63} 
\frac{1}{\xi^2} 
+ \frac{298432}{315} \frac{1}{\xi^3} 
+ \frac{102656}{105} \frac{1}{\xi^4} \right ) 
\ln \left ( \frac{\lambda+1}{\lambda-1} \right ) -\frac{20}{3} 
\frac{1}{\xi^2} 
\ln^2 \left ( \frac{\lambda+1}{\lambda-1} \right ) \right ]~,
\end{eqnarray}
where $T_f$ stands for the colour factor which in QCD is given by 
$T_f=1/2$
(for $C_F$ see below Eq. (\ref{eq11})).
The coefficient function for $F_1$ can be derived from the ones 
obtained
for the structure functions $F_2$ and $F_L$ which are presented in 
\cite{bmsmn}. They are obtained using the same renormalization condition
for the heavy quark loop contribution to the gluon self-energy as given
above.
The result for the unpolarized Bj{\o}rken sum rule is
\begin{eqnarray}
\label{eq14}
&& A_{\rm H}^{F_1,(2)}(Q^2,m^2) =
\left (\frac{\alpha_s(n_f,\mu^2)}{4\pi} \right )^2 C_F T_f
\left [- \frac{2}{105} \xi^2  \ln \xi +
\frac{1}{\lambda^4} \left ( -\frac{4}{105} \xi + \frac{2162}{315} 
+ \frac{30712}{315} \frac{1}{\xi} \right. \right.
\nonumber\\[2ex]
&& \left. \left. + \frac{138848}{315} \frac{1}{\xi^2} 
 + \frac{13696}{21} \frac{1}{\xi^3} \right ) + \frac{1}{\lambda^5} 
\left (
 \frac{2}{105} \xi^2 + \frac{4}{21} \xi - \frac{44}{21} 
- \frac{1000}{21}
\frac{1}{\xi} - \frac{6752}{21} \frac{1}{\xi^2} - \frac{99712}{105} 
\frac{1}{\xi^3} \right. \right.
\nonumber\\[2ex]
&& \left. \left. - \frac{22016}{21} \frac{1}{\xi^4} \right ) 
\ln \left ( \frac{x+1}{x-1} \right ) -\frac{8}{\xi^2} 
\ln^2 \left ( \frac{\lambda+1}{\lambda-1} \right ) \right ],
\quad \mbox{with} \quad \lambda = \sqrt {1 + \frac{4}{\xi}}~.
\end{eqnarray}
Like for the flavour excitation mechanism (Eqs. (\ref{eq8}), 
(\ref{eq9}))
we have checked that the gluon splitting process does not contribute
to the Adler sum rule (\ref{eq7}).
We are also interested in the asymptotic expansions of the expressions 
above.
In the case the quark mass gets much larger than the virtuality of the 
intermediate vector bosons we get
\begin{eqnarray}
\label{eq15}
 A_{\rm H}^{\rm r,(2)}(Q^2,m^2)
& {\raisebox{-2 mm}{$\,\stackrel{=}{{\scriptstyle m^2 \gg Q^2 }}\, $} }&
\left (\frac{\alpha_s(n_f,\mu^2)}{4\pi} \right )^2 C_F T_f \left [
\left ( \frac{16}{45}\xi + \frac{1}{105}\xi^2 \right ) \ln \xi 
- \frac{232}{225}\xi - \frac{1933}{44100}\xi^2 \right. 
\nonumber\\[2ex]
&& \left. - \frac{1}{62370}\xi^4 + \frac{2}{945945}\xi^5 \right ] 
\qquad r=g_1,F_3
\end{eqnarray}
and
\begin{eqnarray}
\label{eq16}
 A_{\rm H}^{\rm F_1,(2)}(Q^2,m^2)
& {\raisebox{-2 mm}{$\,\stackrel{=}{{\scriptstyle m^2 \gg Q^2 }}\, $} }&
\left (\frac{\alpha_s(n_f,\mu^2)}{4\pi} \right )^2 C_F T_f \left [
-\frac{2}{105}\xi^2 \ln \xi - \frac{16}{45}\xi + \frac{1093}{22050}\xi^2
+ \frac{8}{4725}\xi^3  \right.
\nonumber\\[2ex]
&& \left. - \frac{1}{10395}\xi^4 + \frac{8}{945945}\xi^5 \right ]~.
\end{eqnarray}
The expressions show that for infinite mass ($\xi \rightarrow 0$) the 
corresponding heavy flavour
decouples from the radiative correction which is a consequence of the
renormalization condition for the gluon self-energy in the graphs of
Fig. \ref{fig2}. 
\begin{table}
\begin{center}
\begin{tabular}{|c|c|c|c|}\hline
\multicolumn{4}{|c|}{$Q^2=2.5~({\rm GeV/c})^2$} \\ \hline \hline
& $\Delta g_1(Q^2)$ & $\Delta F_1(Q^2)$ & $\Delta F_3(Q^2)$ \\ \hline 
\hline
           &                &                    &             \\
$A(3)$ & 0.795         & 0.847      &  0.797           \\[2mm]
$\delta A^{\rm PMS}(3)$ & $-0.265.10^{-1}$  & $-0.271.10^{-1}$ & 
$-0.265.10^{-1}$   \\[2mm]
$A_c^{(1)}$ &          & $0.169.10^{-1}$  & $-0.541.10^{-2}$  \\[2mm]
$A_c^{(2)}$ &  $-0.688.10^{-3}$ & $-0.199.10^{-3}$  & 
$-0.688.10^{-3}$\\[2mm] 
$A_c^{\rm asymp,(2)}$ & $0.451.10^{-2}$ & $0.406.10^{-2}$ & 
$0.451.10^{-2}$  \\[2mm]
$A_b^{(2)}$ & $-0.131.10^{-3}$  & $-0.253.10^{-4}$ & $-0.131.10^{-3}$\\[2mm]
$A_b^{\rm asymp,(2)}$ & $0.974.10^{-2}$ & $0.755.10^{-2}$ 
& $0.974.10^{-2}$
  \\[2mm] \hline \hline
\multicolumn{4}{|c|}{$Q^2=10~({\rm GeV/c})^2$} \\ \hline \hline
           &                &                    &             \\
$A(3)$ & 0.883         & 0.915      &   0.884          \\[2mm]
$\delta A^{\rm PMS}(3)$ & $-0.586.10^{-2}$  & $-0.600.10^{-2}$ & 
$-0.586.10^{-2}$   \\[2mm]
$A_c^{(1)}$ &    & $0.586.10^{-2}$ & $-0.192.10^{-2}$  \\[2mm]
$A_c^{(2)}$ & $-0.788.10^{-3}$ & $-0.291.10^{-3}$ & $-0.788.10^{-3}$ 
\\[2mm]
$A_c^{\rm asymp,(2)}$ & $0.569.10^{-3}$ & $0.877.10^{-3}$ & 
$0.569.10^{-3}$  \\[2mm]
$A_b^{(2)}$ & $-0.181.10^{-3}$ & $-0.448.10^{-4}$ & $-0.181.10^{-3}$ 
\\[2mm]
$A_b^{\rm asymp,(2)}$ & $0.303.10^{-2}$& $0.252.10^{-2}$ & 
$0.303.10^{-2}$  \\[2mm]  \hline \hline
\multicolumn{4}{|c|}{$Q^2=100~({\rm GeV/c})^2$} \\ \hline \hline
           &                &                    &             \\
$A(3)$ & 0.931     &  0.951     & 0.931   \\[2mm]
$\delta A^{\rm PMS}(3)$ & $-0.121.10^{-2}$ & $-0.124.10^{-2}$ & 
$-0.121.10^{-2}$   \\[2mm]
$A_c^{(1)}$ &        & $0.565.10^{-3}$ & $-0.188.10^{-3}$  \\[2mm]
$A_c^{(2)}$ & $-0.107.10^{-2}$ & $-0.529.10^{-3}$ & $-0.107.10^{-2}$
\\[2mm]
$A_c^{\rm asymp,(2)}$ & $-0.912.10^{-3}$ & $-0.382.10^{-3}$ & 
$-0.912.10^{-3}$  \\[2mm]
$A_b^{(2)}$ & $-0.380.10^{-3}$ & $-0.143.10^{-3}$ & $-0.380.10^{-3}$
\\[2mm]
$A_b^{\rm asymp,(2)}$ & $0.205.10^{-3}$ & $0.363.10^{-3}$ & 
$0.205.10^{-3}$  \\[2mm] \hline
\end{tabular}
\end{center}
\caption{\sf
The contributions to the sum rules originating from the light
quarks $A(n_f)$, Eq. (\ref{eq4}), the charm excitation
$A_c^{(1)}$, Eqs. (\ref{eq10}), (\ref{eq11}), and gluon splitting
into heavy quarks $A_H^{(2)}$ ($H=c,b$), Eqs.~(\ref{eq13}), (\ref{eq14}).
For a comparison we also presented the asymptotic expressions 
$A_H^{\rm asymp,(2)}$, Eq. (\ref{eq19}), in order $\alpha_s^2$.}
\label{tab2}
\end{table}
When the virtuality $Q$ of the vector bosons is much larger than the mass
of
the heavy quark, which implies that the latter behaves like a light
flavour, we obtain
\begin{eqnarray}
\label{eq17}
&& A_{\rm H}^{\rm r,(2)}(Q^2,m^2)
 {\raisebox{-2 mm}{$\,\stackrel{=}{{\scriptstyle Q^2 \gg m^2 }}\, $} }
\left (\frac{\alpha_s(n_f,\mu^2)}{4\pi} \right )^2 C_F T_f \left [
-\frac{20}{3}\frac{1}{\xi^2} \ln^2 \xi - \left ( 4 + \frac{64}{3}
\frac{1}{\xi}
+ \frac{226}{9}\frac{1}{\xi^2} + \frac{32}{5}\frac{1}{\xi^3}  \right. 
\right.
\nonumber\\[2ex]
&& \left. \left.  - \frac{56}{15}\frac{1}{\xi^4} 
+ \frac{256}{63}\frac{1}{\xi^5} \right )
\ln \xi + 8 + \frac{272}{9}\frac{1}{\xi} - \frac{1775}{54}\frac{1}{\xi^2} 
- \frac{536}{75}\frac{1}{\xi^3}
+ \frac{118}{225}\frac{1}{\xi^4} +\frac{7136}{6615}\frac{1}{\xi^5} 
+ \cdots \right ] 
\nonumber\\[2ex]
&& \mbox{with} \qquad r=g_1,F_3 \, \\[2ex]
\label{eq18}
&& A_{\rm H}^{\rm F_1,(2)}(Q^2,m^2)
 {\raisebox{-2 mm}{$\,\stackrel{=}{{\scriptstyle Q^2 \gg m^2 }}\, $} }
\left (\frac{\alpha_s(n_f,\mu^2)}{4\pi} \right )^2 C_F T_f \left [
- \frac{8}{\xi^2} \ln^2 \xi - \left ( \frac{8}{3} + \frac{64}{3}\frac{1}
{\xi}
+ \frac{28}{\xi^2} + \frac{128}{15}\frac{1}{\xi^3}  \right. \right.
\nonumber\\[2ex]
&& \left. \left.  -\frac{16}{3}\frac{1}{\xi^4} 
+ \frac{128}{21}\frac{1}{\xi^5} \right )
\ln \xi + \frac{64}{9} + \frac{320}{9}\frac{1}{\xi} - \frac{35}{\xi^2}
- \frac{1984}{225}\frac{1}{\xi^3} + \frac{4}{9}\frac{1}{\xi^4} 
+\frac{1376}{735}\frac{1}{\xi^5} + \cdots \right ]~.
\end{eqnarray}
The leading terms in the expressions above which are given by the 
constant and the logarithm $\ln \xi$ can be predicted by the 
renormalization
group. 
The general form up to order $\alpha_s^3$ becomes
\begin{eqnarray}
\label{eq19}
A_{\rm H}^{\rm r, asymp}(Q^2,m^2)&=& \left (
\frac{\alpha_s(n_f,\mu^2)}{\pi} \right )^2
\left \{ \frac{1}{6} a_1 \ln \left ( \frac{Q^2}{m^2} \right ) + b_2 
\right \}
+ \left (\frac{\alpha_s(n_f,\mu^2)}{\pi} \right )^3 \left \{ a_1 \left ( 
-\frac{8}{9} \right. \right. 
\nonumber\\[2ex]
&& \left. \left. + \frac{1}{18}n_f \right ) \ln^2 \left ( \frac{Q^2}{m^2} 
\right ) +a_1 \left ( \frac{11}{12} - \frac{1}{18} n_f \right )
\ln \left ( \frac{Q^2}{m^2} \right ) \ln \left ( \frac{\mu^2}{m^2} \right )
 + \left [\frac{19}{24} a_1 
\right. \right.
\nonumber\\[2ex]
&& \left. \left. + \frac{1}{3} \Big (a_2 + b_2 n_f \Big )
- b_2 \left (\frac{11}{2} - \frac{1}{3}(n_f+1) \right )
\right ]\ln \left ( \frac{Q^2}{m^2} \right ) \right.
\nonumber\\[2ex]
&& \left. + b_2 \left ( \frac{11}{2} - \frac{1}{3}n_f \right ) \ln \left 
( \frac{\mu^2}{m^2} \right ) + b_3 + c_3 \Big (2 n_f + 1 \Big )   
\right \}~,
\end{eqnarray}
where the coefficients $a_i,b_i,c_i$ originate from the light quark
contributions presented in table \ref{tab1}. Notice that in the 
equation
above we have already substituted the values of coefficients $\beta_0, 
\beta_1$,
Eq. (\ref{eq5}), appearing in the series expansion of the
$\beta$-function.
The large logarithms of the type $\ln Q^2/m^2$, which originate from all
heavy quark loop insertions like in Fig. \ref{fig2}, can be absorbed into
the running coupling constant. This is equivalent to a redefinition given by
\begin{eqnarray}
\label{eq20}
\alpha_s(n_f,\mu^2)&=&\alpha_s(n_f+1,\mu^2) \left [ 1
+ \frac{\alpha_s(n_f+1,\mu^2)}{\pi} \left ( \beta_0(n_f+1)-\beta_0(n_f) 
\right )
\ln \left ( \frac{\mu^2}{m^2} \right ) \right.
\nonumber\\[2ex]
&& \left. + \left (\frac{\alpha_s(n_f+1,\mu^2)}{\pi} \right )^2
\left\{   \left ( \beta_0(n_f+1)-\beta_0(n_f) \right )^2
\ln^2 \left ( \frac{\mu^2}{m^2} \right ) \right. \right.
\nonumber\\[2ex]
&& \left. \left. +  \left ( \beta_1(n_f+1)-\beta_1(n_f) \right )
\ln \left ( \frac{\mu^2}{m^2} \right ) \right \} \right ]~.
\end{eqnarray}
If we add the asymptotic expression (\ref{eq19}) for the heavy quark loop 
contributions to the perturbation series for the light quarks in 
Eq. (\ref{eq4}) we obtain after substitution of $\alpha_s(n_f,\mu^2)$ the 
following result
\begin{eqnarray}
\label{eq21}
A_{\rm H}^{\rm r, asymp}(Q^2,m^2)+ A^{\rm r}(n_f,Q^2)
=A^{\rm r}(n_f+1,Q^2)~.
\end{eqnarray}
Therefore for $Q^2 \gg m^2$ we obtain the expression of the perturbation
series for massless flavours again but wherein now 
the number of light flavours is enhanced by one unit. The question is
at which $Q^2$ this will happen. Here we will give the answer for
the charm quark because the deep inelastic sum rules are studied in the
region $2 < Q^2 < 100 ({\rm GeV/c})^2$. 

In our analysis we will use the three-loop corrected running coupling
constant which satisfies the matching conditions \cite{marc}
\begin{eqnarray}
\label{eq22}
\alpha_s \left (n_f,\Lambda_{n_f},\mu^2\right) =
\alpha_s \left (n_f+1,\Lambda_{n_f+1},\mu^2\right) \quad \mbox{at} \quad
\mu=m_{n_f}~.
\end{eqnarray}
If we choose $\Lambda_3= 397~{\rm MeV/c}$ (${\overline {\rm MS}}$-scheme) 
we get $\alpha_s(3,\mu_0^2)=0.375$ for $\mu_0^2=2.5~({\rm GeV/c})^2$. 
These values were obtained from a comparison of the polarized Bj{\o}rken
sum rule with the data carried out in \cite{elka}. Following the matching
conditions in Eq. (\ref{eq22}) one gets $\alpha_s(5,M_Z^2)=0.122$ (here
$\Lambda_5= 259~{\rm MeV/c}$) which lies a little bit above the world 
average of $\alpha_s(5,M_Z^2)=0.119$. Nevertheless we will use 
$\alpha_s(3,\mu_0^2)$ as a starting point.
The values for the heavy flavour
masses are chosen to be $m_c=1.5~{\rm GeV/c^2}$, $m_b=4.5~{\rm GeV/c^2}$
and  $m_t=173.8~{\rm GeV/c^2}$. Further the number of light flavours in 
$A^{\rm r}$, Eq. (\ref{eq4}), and the running coupling constant is taken 
to be $n_f=3$ irrespective of the value of $Q^2$. In table \ref{tab2} we 
have presented for $Q^2=2.5, 10, 100~{\rm GeV/c^2}$ the light and heavy flavour 
contributions to the perturbation series denoted by $A(3)$ and $A_H^{(i)}$ 
($H=c,b$), respectively. In the case of $A(3)$ , Eq. (\ref{eq4}), we only 
consider the exact perturbation series corrected up to order $\alpha_s^3$ 
and omitted the fourth order estimate $\delta A^{\rm PMS}(3)$, 
Eq. (\ref{eq6}), which is listed separately in table \ref{tab2}. 
The charm contribution, represented by the order $\alpha_s$ corrected quantity 
$A_c^{(1)}$, is given by Eqs. (\ref{eq10}), (\ref{eq11})
which does not appear in the case of the polarized Bj{\o}rken sum rule.
The remaining heavy flavour contributions show up in order $\alpha_s^2$
and they are represented in the table by $A_c^{(2)}$ , $A_b^{(2)}$ (see Eqs.
(\ref{eq13}), (\ref{eq14})). The top quark contribution is so small that it 
is neglected. Besides the exact results for $A_H^{(2)}$ we have also made a 
comparison with the asymptotic expression given by the order $\alpha_s^2$ 
contribution $A_H^{{\rm asymp},(2)}$ (see Eq. (\ref{eq19})) derived in the 
limit $Q^2 \gg m^2$. From table \ref{tab2} we infer that the
heavy flavour contributions are rather small even when compared with the
estimate $\delta A^{\rm PMS}(3)$. Only in the case of $\Delta F_1(Q^2)$
at $Q^2=2.5~{\rm GeV/c^2}$ the charm component is of the same size as the
order $\alpha_s^2$ estimate and it amounts to 0.017 which is about 2\% 
of the light quark contribution given by $A^{F_1}(3)=0.847$. 
This effect can be wholly attributed
to the charm excitation mechanism (see Eqs. (\ref{eq8}), (\ref{eq9})) 
represented by $A_c^{(1)}$, which also dominates $\Delta F_3(Q^2)$.
At larger values of $Q^2$, $A_c^{(1)}$ decreases and it becomes of the same 
order of magnitude as $A_c^{(2)}$ which is due to the gluon splitting 
mechanism. Notice that the bottom quark contribution is always smaller than 
the charm quark component. The behaviour of the heavy quark contributions 
follows from their asymptotic behaviour at small and at large $Q^2$,
see e.g. Eqs. (\ref{eq15})-(\ref{eq18}). At increasing $Q^2$ the charm 
excitation contribution $A_c^{(1)}$ is decreasing whereas the gluon splitting 
part $A_c^{(2)}$ becomes larger. At $Q^2=100~{\rm GeV/c^2}$, which is
about $Q=6.5~m$, the latter gets 
closer to its asymptotic expression $A_c^{{\rm asymp},(2)}$. However, there is
still a discrepancy between the exact and asymptotic expressions which 
in the case of the sum rules $\Delta g_1(Q^2)$ and $\Delta F_3(Q^2)$ amounts 
to 15 \%.  For $\Delta F_1(Q^2)$ this is much worse and the difference between
the exact and asymptotic expression
is 28 \% w.r.t. the exact one. In the case of the bottom quark one
needs much larger values before $A_b^{(2)} \sim A_b^{{\rm asymp},(2)}$ 
which occurs for $Q^2 > 1000~{\rm GeV/c^2}$. 
In order to get $A_H^{{\rm asymp},(2)}=A_H^{(2)}$ within 1 \% one needs
the value $Q > 25~m$. Hence we can conclude that
the large logarithmic terms given by $\ln(Q^2/m^2)$ start to dominated
the heavy flavour contribution for $Q > 6.5~m$ which means that from 
this value
onwards the heavy flavour behaves like a light quark. Therefore only 
for $Q > 6.5~m$ the large logarithms have to be resummed
as is explained below Eq. (\ref{eq20}) which will lead to a $n_f+1$ 
flavour
description. This means that the matching condition $\mu=m_{n_f}$
has to be changed into $\mu = 6.5~m_{n_f}$. Using this new matching
condition we 
get, starting from $\alpha_s(3,2.5)=0.375$  as our experimental input 
value, 
the result $\alpha_s(5,M_Z^2)=0.114$. The latter is very close to the 
value
obtained in fixed target deep inelastic scattering experiments given by
$\alpha_s(5,M_Z^2)=0.113$ \cite{vimi}. However, from the analysis above
we think that the matching conditions as presented in Eq. (\ref{eq22})
are rather artificial. There is no specific scale where nature suddenly
jumps from an $n_f$-flavour to an $n_f+1$-flavour scheme. Moreover the
relevant scale in deep inelastic scattering is $q^2=-Q^2$ which is 
spacelike
rather than $p^2 \ge 4 \cdot m_{n_f}^2$ which is timelike. 
Here $p$ denotes the
gluon momentum in the graphs of Fig. \ref{fig2}. Therefore, in principle
all heavy flavour channels may contribute for spacelike processes which 
proceeds via the coefficient functions rather than through the running 
coupling constant. The decoupling of the heavy flavours from the 
perturbation series is then ruled by the Appelquist--Carazzone theorem 
\cite{apca}. In order to get more continuity between the large and small 
$Q^2$ regions we substitute on the l.h.s. of Eq. (\ref{eq21}) at $n_f=3$ 
the coupling constant by 
\begin{eqnarray}
\label{eq23}
\alpha_s(\mu^2,3)=\alpha_s^{\rm MOM}(\mu^2) \left [ 1 + 
\frac{\alpha_s^{\rm MOM}(\mu^2)}{4\pi} U_1 +
\left (\frac{\alpha_s^{\rm MOM}(\mu^2)}{4\pi}\right )^2 \Big ( U_2 +U_1^2
\Big ) + \cdots \right ] \,,
\end{eqnarray}
with
\begin{eqnarray}
\label{eq24}
U_i = T_f \sum_{n_f=4}^6 \left [ \Pi_i \left (\frac{\mu^2}{m_{n_f}^2}
\right ) -
\Pi_i \left (\frac{\mu_0^2}{m_{n_f}^2}\right ) \right ] \quad \mbox{with}
\quad m_4=m_c, \, m_5=m_b, \, m_6=m_t\,.
\end{eqnarray}
Here we are starting from a low input scale $\mu_0$ so that one is sure
that at this value the perturbation series is described by a 
three-flavour number scheme.
The functions $\Pi_i$ are the order $\alpha_s^i$ contributions to the
gluon self-energy which can be attributed to the heavy quark loop only.
They
are presented in \cite{jeta} up to order $\alpha_s^2$. For
$\mu^2 \gg m_{n_f}^2$ and $m_{n_f}^2 \gg \mu_0^2$ the expression in 
Eq. (\ref{eq23}) tends to its asymptotic result in Eq. (\ref{eq20}) 
provided 
one considers in the sum of Eq. (\ref{eq24}) one flavour only. 
Finally one can resum the self-energies so that our new running coupling
constant becomes
\begin{eqnarray}
\label{eq25}
\alpha_s^{\rm MOM}(\mu^2)=\frac{\alpha_s(\mu^2,3)}{1+ 
\frac{\alpha_s(\mu^2,3)}{4\pi} U_1 +
\frac{\alpha_s(\mu^2,3)}{4\pi} \Big (U_2/U_1 \Big ) \ln \left ( 1 +
\frac{\alpha_s(\mu^2,3)}{4\pi} U_1 \right ) }~.
\end{eqnarray}
The coupling constant above was proposed in the context of the momentum 
subtraction (MOM) scheme in \cite{shir} (for earlier work on this subject 
see also \cite{yoha}) where it also includes
the resummation of the light quark contributions which we did not perform
 in this paper.
The procedure outlined above guarantees that one gets a smooth transition 
between the low and large $Q^2$ regions. When e.g. $\mu^2=Q^2$ and 
$Q^2 \gg m_{n_f}^2$ the large logarithms $\ln(Q^2/m_{n_f}^2)$ occurring in the
functions $U_i$ (\ref{eq24}) cancel against the corresponding terms
in the perturbations series of the heavy quark component 
$A_H(Q^2,m_{n_f}^2)$
of which the asymptotic expression is given by Eq. (\ref{eq19}). In this 
way 
one gets effectively a $(n_f+1)$-flavour description. If
$Q^2 \ll m_{n_f}^2$ then
$U_i \sim 0$ and no large corrections appear in $A_H(Q^2,m_{n_f}^2)$
(decoupling of heavy quarks !) so that
one gets the $n_f$-flavour representation for the whole perturbation 
series.
Following our approach and choosing $\mu_0^2=2.5~{\rm GeV/c^2}$ in
Eq. (\ref{eq24}) we get $\alpha_s(5,M_Z^2)=0.117$ which is closer to the
LEP measurement. Finally we checked that the numbers in the table hardly 
change in passing from the $n_f=3$ ${\overline {\rm MS}}$-scheme to the 
MOM-scheme. The maximal deviation is observed for $\Delta g_1(Q^2)$ where
at $Q^2=100 ~{\rm GeV/c^2}$ the latter scheme leads to a decrease of
$A^{g_1}(3)$ by 
0.002 so that for this value of $Q^2$ the resummation effect is very small.
Summarizing our findings we have calculated the heavy quark contribution
to several deep inelastic sum rules. The corrections to the three-loop
corrected perturbation series, computed for light quarks, are very small.
Only at low $Q^2$ the correction in the case of the unpolarized 
Bj{\o}rken
sum rule $\Delta F_1(Q^2)$ is noticeable where it is of the same order
of magnitude as the order $\alpha_s^4$-estimate. The quark component
of the sum rule attains its asymptotic value at much larger scales as 
given by the usual matching conditions. Matching at larger scales i.e.
$\mu=6.5~m_{n_f}$ leads to a smaller value of the 
running coupling constant at the $Z$-boson mass. The unnatural matching
conditions which are characteristic of the ${\overline {\rm MS}}$-scheme
can be replaced by expressing the perturbation series in the MOM-scheme. 

\vspace{2mm}
\noindent
{\bf Acknowledgements.}
The authors would like to thank V. Ravindran, F. Jegerlehner and A.~Kataev
for useful discussions. This work is supported by the EC-network under
contract FMRX-CT98-0194.


\begin{thebibliography}{99}
\bibitem{ckt}
K.G. Chetyrkin, A.L. Kataev and F.V. Tkachov, Nucl. Phys. {\bf B174} 
(1980)
345.
\bibitem{chtk}
K.G. Chetyrkin and F.V. Tkachov, Nucl. Phys. {\bf B192} (1981) 159.
\bibitem{chsm}
K.G. Chetyrkin and F.V. Tkachov, Phys. Lett. {\bf B114} (1982) 340;\\
K.G. Chetyrkin and V.A. Smirnov, Phys. Lett. {\bf B144} (1984) 419.
\bibitem{form}
A. Vermaseren, "Symbolic Manipulation with FORM", Published by CAN,
Amsterdam, The Netherlands, ISBN 90-74116-01-9.
\bibitem{bjork1}
J.D. Bj{\o}rken, Phys. Rev. {\bf 148} (1966) 1467, Phys. Rev. {\bf D1} 
(1970) 1376.
\bibitem{bjork2}
J.D. Bj{\o}rken, Phys. Rev. {\bf 163} (1967) 1767.
\bibitem{grls}
J.J.Gross and C.H. Llewellyn Smith, Nucl. Phys. {\bf B14} (1969) 337.
\bibitem{kast}
A.L. Kataev and V.V. Starshenko, Mod. Phys. Lett. {\bf A10} (1995) 235.
\bibitem{stev}
P.M. Stevenson, Phys. Rev. {\bf D23} (1981) 2916.
\bibitem{grun}
G. Grunberg, Phys. Lett. {\bf B221} (1980) 70, Phys. Rev. {\bf D29} 
(1984) 2315.
\bibitem{sael}
M.A. Samuel, J. Ellis, M. Karliner, Phys. Rev. Lett. {\bf 74} (1995) 
4380.
\bibitem{elga}
J. Ellis, E. Gardi, M. Karliner and M.A. Samuel, Phys. Lett. {\bf B366}
(1996) 268.
\bibitem{elka}
J. Ellis and M. Karliner, Phys. Lett. {\bf B341} (1995) 397.
\bibitem{latk}
S.A. Larin, F.V. Tkachev, J.A.M. Vermaseren, Phys. Rev. Lett. {\bf 66} 
(1991) 862.
\bibitem{lave}
S.A. Larin and J.A.M. Vermaseren, Phys. Lett. {\bf B259} (1991) 345.
\bibitem{adl}
S.L. Adler, Phys. Rev. {\bf 143} (1966) 1144.
\bibitem{gott}
Th. Gottschalk, Phys. Rev. {\bf D23} (1981) 56.
\bibitem{gkr}
M. Gl\"uck, S. Kretzer and E. Reya, Phys. Lett. {\bf B380} (1996) 171.
\bibitem{teve}
O.V. Teryaev and O.L. Veretin, preprint hep-ph/9602362.
\bibitem{bmsn}
M. Buza, Y. Matiounine, J. Smith, W.L. van Neerven, Nucl. Phys. 
{\bf B485}
(1997) 420.
\bibitem{bmsmn}
M. Buza, Y. Matiounine, J. Smith, R. Migneron and W.L. van Neerven, 
Nucl. Phys. {\bf B472} (1996) 611.
\bibitem{marc}
W.J. Marciano, Phys. Rev. {\bf D29} (1984) 580.
\bibitem{vimi}
M. Virchaux and A. Milsztajn, Phys. Lett. {\bf B274} (1992) 221.
\bibitem{apca}
T. Appelquist and J. Carazzone, Phys. Rev. {\bf D11} (1975) 2856.
\bibitem{jeta}
F. Jegerlehner and O.V. Tarasov, DESY 98/093, hep-ph/9809485.
\bibitem{shir}
D.V. Shirkov, Teor. Mat. Fiz. {\bf 93} (1992) 466, Nucl. Phys. {\bf B371}
(1992) 467.
\bibitem{yoha}
T. Yoshino and K. Hagiwara, Z. Phys. {\bf C24} (1984) 185.
\end{thebibliography}
\end{document}